\documentclass[twocolumn,showpacs,preprintnumbers,amsmath,amssymb,superscriptaddress]{revtex4-1}
\usepackage{graphicx}% Include figure files
\usepackage{dcolumn}% Align table columns on decimal point
\usepackage{bm}% math
\usepackage{color}% color
\usepackage{pifont}
\usepackage{natbib}
\usepackage{hyperref}
\hypersetup{
colorlinks = true,
urlcolor   = blue,
linkcolor  = blue,
citecolor  = blue
}

\begin{document}

\preprint{APS/PRB}
\title{Charge Order Stabilized Quantum Spin Liquid in Hollandite K$_2$V$_8$O$_{16}$}

\author{Ola~Kenji~Forslund}
 \email{okfo@kth.se}
\author{Elisabetta~Nocerino}
\affiliation{Department of Applied Physics, KTH Royal Institute of Technology, SE-106 91 Stockholm, Sweden}
\author{Masahiko~Isobe}
\affiliation{Max Planck Institute for Solid State Research, Heisenbergstra\ss e 1, 70569 Stuttgart, Germany}
\author{Daniel~Andreica}
\affiliation{Faculty of Physics, Babes-Bolyai University, 400084 Cluj-Napoca, Romania}
\author{Stephen~Cottrell}
\affiliation{ISIS Facility, Rutherford Appleton Laboratory, Chilton, Didcot Oxon OX11 0QX, United Kingdom}
\author{Hidenori~Takagi}
\affiliation{Max Planck Institute for Solid State Research, Heisenbergstra\ss e 1, 70569 Stuttgart, Germany}
\author{Yasmine~Sassa}
\affiliation{Department of Physics, Chalmers University of Technology, SE-41296 G\"oteborg, Sweden}
\author{Jun~Sugiyama}
\affiliation{Neutron Science and Technology Center, 
Comprehensive Research Organization for Science and Society (CROSS),Tokai, Ibaraki 319-1106, Japan}
\author{Martin~M\aa nsson}
\email{condmat@kth.se}
\affiliation{Department of Applied Physics, KTH Royal Institute of Technology, SE-106 91 Stockholm, Sweden}
\date{\today}

\begin{abstract}
%Heisenberg spin-1/2 chains are the archetypical one dimensional model for describing the detailed ground state in a quantum matter. The developed models can be experimentally verified through measurements of correlations functions.
Quantum spin liquid is an elusive state that display strong many-body entanglement with potential applications in future quantum computing. This study reports muon spin relaxation ($\mu^+$SR) measurements on a novel high-pressure synthesized material, the Hollandite K$_{2}$V$_8$O$_{16}$. In this quasi-one-dimensional compound, charge ordering (CO) at $T_{\rm MIT}\approx160$~K effectively isolates half of the vanadium chains and model-like Heisenberg spin-1/2 chains are realized. Our zero field $\mu^+$SR measurements show exponential like relaxation down to the lowest temperature $T=100$~mK and the absence of long range ordering is confirmed. The relaxation rate is found to be temperature independent below $T_{\rm QSL}\approx2$~K and measurements in longitudinal field confirms a highly dynamic ground state. These results represents the first confirmation of quantum spin liquid (QSL) behavior within the Hollandite family, stabilized by the CO. Finally, the presence of strong local electron correlation and one dimensional Fermi surface suggest this QSL to be a gapless Tomonaga-Luttinger liquid (TLL), which here uniquely presents itself in a stoichiometric compound under zero applied magnetic field and at ambient pressure.
\end{abstract}

\keywords{short keywords that describes your article}

\maketitle

%\section{\label{sec:Intro}Introduction}
Low dimensional magnetism is a field that has developed tremendously over the last decades. From theoretical point of view, exactly solvable solution were more easily obtained without the need to consider the complicated models in 3D \cite{Bethe1931}. It was later realised that these model were not only purely theoretical but some compound actually exhibit signatures of low dimensional magnetism. This has driven low dimensional spin systems to be an ideal playground in order to study and find new states of matter and general concepts of many body physics, given that quantum fluctuations are more prominent in low dimensional systems \cite{Parkinson2010, Diep2013}. 

%Beyond the traditionally thermally stabilised long range magnetic orders, a wealth of phases can be generated via an interplay of orbital, spin and charge degrees of freedom in lower dimensions: Luttinger liquid \cite{Haldane1981}, mott insulator \cite{MOTT1968}, spin liquids \cite{Zhou2017}, spin-Peierls \cite{Shaz2005} states are just few of many available \cite{Schollwock2008}. Spin-Peierls transition was shown in CuGeO \cite{Hase1993}. 

One of the more elusive quantum states is the so-called quantum spin liquid (QSL) that was first proposed by Anderson \cite{Anderson_1973}. The QSL embodies strong entanglement that generates a virtual smörgåsbord of unique physical properties that are interesting for both fundamental understanding \cite{QSL_Review} as well as future applications in quantum computing and spintronics \cite{Quantum_Computing}.
%Previously, the research on QSL states were mainly theoretically focused, however, few recent model materials have allowed also experimental evidence and systematic studies. 
In general, QSL may arise from geometrical frustration in higher dimensions and result in resonance valance bonds liquids or possible Kitaev quantum spin liquid \cite{KQSL_Review} in honeycomb lattices \cite{Honeycomb}. For 1D systems however, spinon excitation were clarified for antiferromagnetic spin-1/2 chains \cite{Faddeev1981, Takhtajan1982} while Haldane \cite{Haldane1983_1, Haldane1983} discovered the fundamental difference between integer and half integer spin chains. Realisation of Spin-1 chains include CsNiF$_3$\cite{Ramirez1982}, which demonstrated magnetic solitons. KCuF$_3$ \cite{Tennant1993} and Sr$_2$CuO$_3$ \cite{Zaliznyak2004} on the other hand have shown to be a realisation of an antiferromagnetic spin-1/2 chains for which inelastic neutron scattering confirmed a QSL via the presence of spinon excitations. 

% 1 col. figure
\begin{figure*}[ht]
  \begin{center}
    \includegraphics[keepaspectratio=true,width=\textwidth]{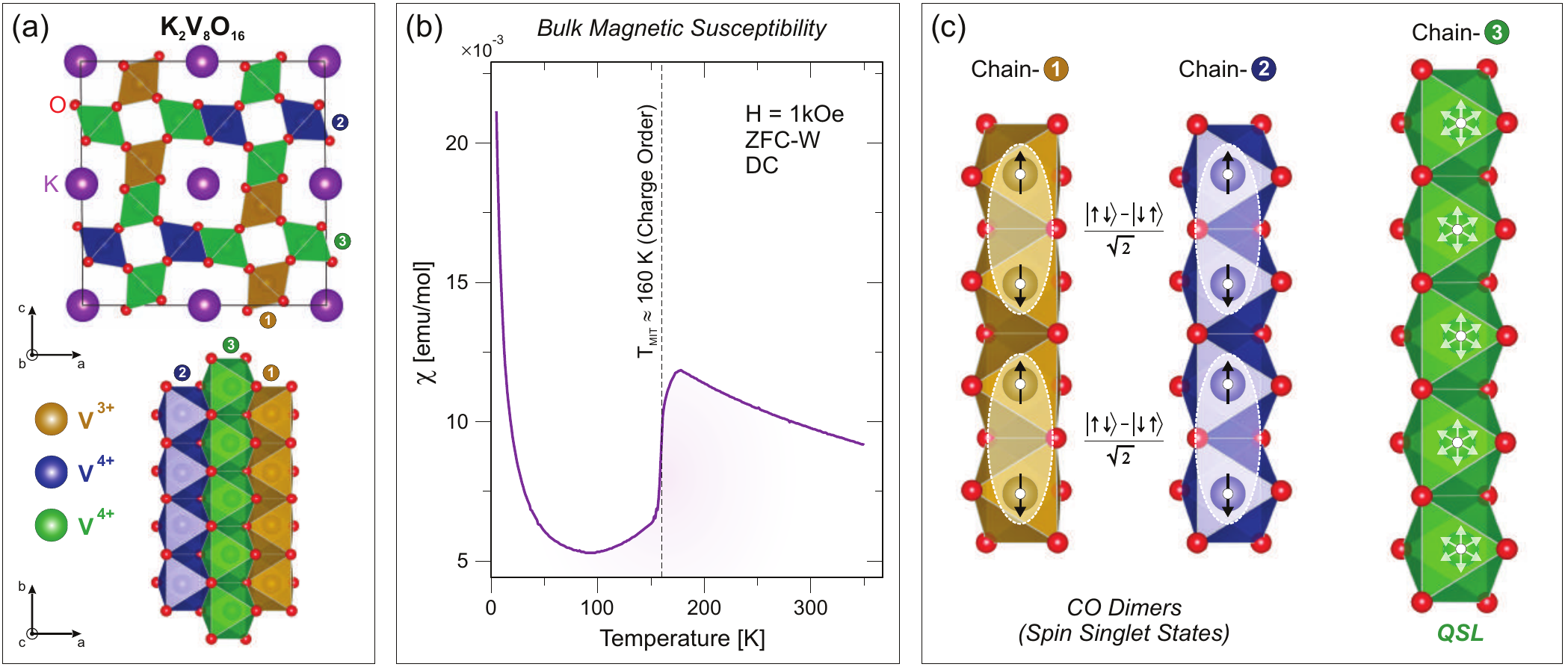}
  \end{center}
  \caption{(a) The CO pattern revealed by XRD measurements presented in Ref.~\onlinecite{Komarek2011}. It is composed of dimerized V$^{3+}$ (orange, \textcircled{\small{1}}) and V$^{4+}$ (blue, \textcircled{\small{2}}) chains and a non-dimerzied V$^{4+}$ (green, \textcircled{\small{3}}) chains. (b) DC-Magnetic susceptibility measured in ZFC protocol with $H = 1$~kOe revealing the 50\% decrease with the CO formation at $T_{\rm MIT}\approx160$~K \cite{Isobe2006, Isobe2008}. (c) Combining the XRD results \cite{Komarek2011} with theoretical predictions of spin singlet formation suggest that the \textcircled{\small{3}} chains are effectively isolated spin-1/2 chains [see also panel (a)], which are available to form the QSL state (as confirmed by the current $\mu^+$SR investigation).}
  \label{fig:Crystal}
\end{figure*}

In search for novel physical properties and phases, the use of high-pressure synthesis is a effective route for stabilizing otherwise inaccessible crystal structures and materials. Among such compounds we find K$_2$V$_8$O$_{16}$, which belongs to the Hollandite family. This group of materials has attracted a lot of attention during the recent years for their intriguing electronic and magnetic properties, both under pressure \cite{Yamauchi2011, Yamauchi2015, Forslund2017, Forslund2019} and ambient conditions \cite{Shimizu2011, Sugiyama2012, Ohta2012, Toriyama2012, Bhobe2015, Kim2016}. Hollandites can be described by a general chemical formula A$_x$M$_8$O$_{16}$ (A = alkali/alkaline-earth metal, x = 1-2, and M = transition metal) with edge shared M$_2$O$_6$ octahedra zigzag double chain structure [see Fig.~\ref{fig:Crystal}(a)]. The chains run along the crystallographic c-axis and are connected via corner shared oxygen atoms and with the A cation at the tunnel site \cite{Tamada1996}. K$_2$V$_8$O$_{16}$ undergoes $T_{\rm MIT}=160$~K, driven by a charge order (CO) formation \cite{Isobe2006, Horiuchi2008}, which is also visible in the bulk magnetic susceptibility that display a sudden and strong reduction (but not to zero!) at $T_{\rm MIT}=160$~K [Fig.~\ref{fig:Crystal}(b)]. Previous zero field (ZF) muon spin relaxation ($\mu^+$SR) measurement suggested that this CO formation is accompanied with imperfect spin singlet formation, $i.e.$ some V spins are still unpaired even at lower temperatures (2~K) \cite{Chow2012}. Indeed, single crystal X-ray diffraction (XRD) \cite{Komarek2011} observed a dimerization in half of the V chains [chains \textcircled{\small{1}} and \textcircled{\small{2}} in Fig.~\ref{fig:Crystal}(a,c)], causing a 50\% decrease in the magnetic susceptibility [Fig.~\ref{fig:Crystal}(b)] due to the formation of spin singlets \cite{Isobe2006, Komarek2011, Kim2016}, followed by a structural change from a tetragonal to a monoclinic symmetry. Consequently, the remaining chains [\textcircled{\small{3}} in Fig.~\ref{fig:Crystal}(a,c)] can be considered as isolated spin$-1/2$ Heisenberg chains, sustained within a 'sea' of spin-dimerized chains \cite{Kim2016}.

In order to elucidate the detailed properties and behavior of the proposed isolated \textcircled{\small{3}}-chains we have conducted a muon spin relaxation ($\mu^+$SR) study of the K$_2$V$_8$O$_{16}$ compound down to $100$~mK. $\mu^+$SR is a very sensitive technique for both long- and short-range correlations in both static and dynamical forms. Clearly, $\mu^+$SR is a local probe and lacks the detailed Q-resolution of neutron scattering. However, the muon's larger gyromagnetic ratio gives it a uniquely high sensitivity to even small magnetic moments using small sample volumes. Further, similar to neutron scattering, $\mu^+$SR is also able to conduct investigations at both ultra-low temperatures as well as under zero-applied magnetic fields. Hence, $\mu^+$SR is an optimal technique to investigate subtle spin properties (like the QSL) also in powder samples available in small amounts yielded by high-pressure synthesis.

%Skriv om "mer positivt, high-p synthesis -> muSR som likt INS kan mäta under ZF och låga temperaturer... Kapa summeringen till
%Naturally, direct observation of Spinon excitation is a strong evidence for the formation of quantum spin liquid. In certain cases however, the amount of sample available or the size of the crystal makes inelastic neutron scattering not feasible. Therefore, we have initiated muon spin relaxation ($\mu^+$SR) measurements on K$_2$V$_8$O$_{16}$ down to $T=100$~mK, a sample synthesized under high pressure and temperature. $\mu^+$SR is sensitive to both static and dynamic internal field fluctuations, and supporting evidence of QSL can be obtained even with data from powder samples of very small mass. 

% MOVE TO CONCLUSIONS ???
%======================================================================
%In detail, absence of long range magnetic order is confirmed by the exponential like relaxation at temperatures down to $T=100$~mK. Moreover, longitudinal field (LF) measurements confirm the origin of these temperature independent relaxation to be dynamic and supports the scenario of a QSL. These results represent the first instance of a QSL in the hollandite family, which together with the 1D nature and electron correlation present makes K$_2$V$_8$O$_{16}$ a prime candidate for a Tomonaga-Luttinger liquid (TLL). 
%======================================================================

%\section{\label{sec:exp}Experimental Setup}
The current powder samples of K$_2$V$_8$O$_{16}$ were synthesized through a high-pressure, high-temperature, solid state reaction of KVO$_3$, V$_2$O$_3$ and V$_2$O$_5$, which is described in greater detail in Ref.~\onlinecite{Isobe2006}. The $\mu^+$SR measurements were carried out at the MUSR surface muon beamline at ISIS pulsed muon source, UK. Approximately 600~mg of sample was mounted onto a silver holder and a dilution fridge cryostat was utilized to reach temperatures down to $100$~mK. The $\mu^+$SR data has been analysed with the \texttt{musrfit} \cite{musrfit} software suite.

\begin{figure}[ht]
  \begin{center}
    \includegraphics[keepaspectratio=true,width = 80 mm]{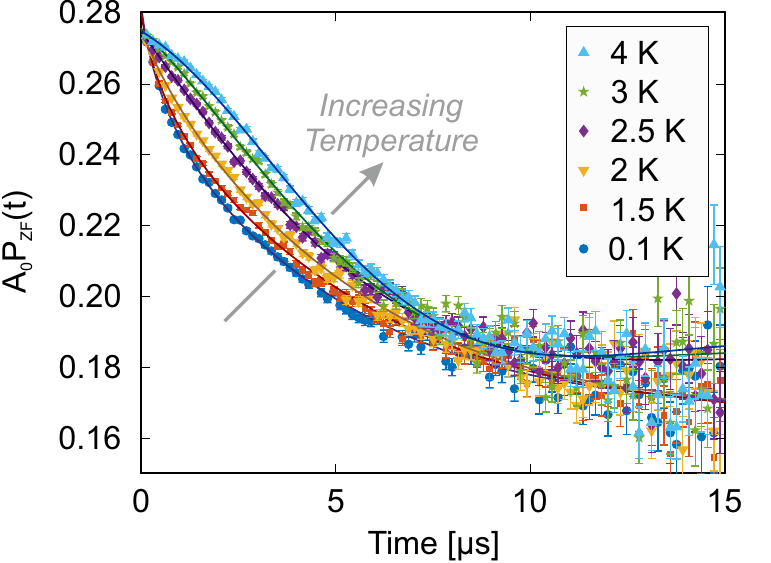}
  \end{center}
  \caption{Zero field (ZF) time spectra at selected temperatures. The solid lines are the best fits obtained using Eq.~(\ref{eq:ZF_high}) and Eq.~(\ref{eq:ZF_low}), depending on the temperature.}
  \label{fig:ZFSpec}
\end{figure}

%\section{\label{sec:results}Results}
Zero field (ZF) $\mu^+$SR time spectra for selected temperatures are displayed in Fig.~\ref{fig:ZFSpec}. At 4~K, an exponentially damped Kubo-Toyabe (KT) depolarisation is observed, which become more exponential-like at lower temperatures. Therefore, the fitting function was divided into two temperature regions: for $T> 2$~K, the measured ZF time spectrum was fitted with an exponentially relaxing static KT together with a background component to account for the sample holder, whereas a stretched exponential together with the same sample holder contribution was fitted for $T\leq2~$K:

\begin{eqnarray}
 A_0 \, P_{\rm ZF}(t) = A_{\rm KT}G^{\rm SGKT}(\Delta_{\rm KT},t)e^{-\lambda_{\rm KT}t}\cr +A_{\rm BG}e^{-\lambda_{\rm BG}t}>2~{\rm K},
\label{eq:ZF_high}
\end{eqnarray}

\begin{eqnarray}
 A_0 \, P_{\rm ZF}(t) = A_{\rm S}e^{-(\lambda_{\rm S}t)^{\beta}}+A_{\rm BG}e^{-\lambda_{\rm BG}t}\leq2~{\rm K},
\label{eq:ZF_low}
\end{eqnarray}

where $A_0$ is the initial $t=0$ asymmetry that depends on the detector geometry of the instrument and $P_{\rm ZF}(t)$ represents the muon polarisation function under ZF configuration. $A_{\rm KT}$ and $\lambda_{\rm KT}$ are the respective asymmetries and the associated depolarisation rate. The asymmetries represent the magnetic volumic fraction, $i.e.$ the fraction of muon experiencing that particular field distribution descried by the associated polarisation function. $G^{\rm SGKT}$ represents a static Gaussian-KT and has the form

\begin{eqnarray}
P_{\rm SGKT}(t) = \frac{1}{3}+\frac{2}{3}(1-t^2\Delta^2)e^{-\frac{t^2\Delta^2}{2}},
\label{eq:GKT}
\end{eqnarray}

which is derived assuming isotropic randomly Gaussian-distributed magnetic moments with a field distribution width $\Delta_{\rm KT}$. The ZF $\mu^+$SR time spectrum at $4$~K was found to fully decouple (not shown) even for small values of externally applied longitudinal fields, LF = 5-20 G (here, LF refers to a magnetic field direction parallel to the initial muon spin polarisation). Therefore, the isotropically distributed Gaussian field distribution at higher temperatures is understood to be originated from static nuclear moments, composed mostly of $I_{^{51}\rm V}=7/2$. The exponential ($\lambda_{\rm KT}$) on the other hand accounts for fluctuations stemming from localised V$^{4+}$ electron moments, that depolarises the muon spin polarisation on top of the nuclear depolarisation. 

$A_{\rm X}$  and $\lambda_{\rm X}$ are the respective asymmetry and relaxation rates where the index indicates if it is from the sample ($X=$~S) or the sample holder ($X=$~BG). Indeed, very low values of $\lambda_{\rm BG}=0.0056(5)\mu$s$^{-1}$ are obtained at base temperature, as expected for Ag contribution. Moreover, since $A_{\rm BG}$ is expected to be temperature independent, the value was fixed to $A_{\rm BG}=0.18462$ for the whole temperature range, a value estimated from the 4~K measurement for which the sample polarisation function is known to be exponential KT function. $\beta$ on the other hand is the stretched exponent where values $<1$ implies the presence of distributions of relaxation rates \cite{Forslund2021_La}. 

\begin{figure}[ht]
  \begin{center}
    \includegraphics[keepaspectratio=true,width = 80 mm]{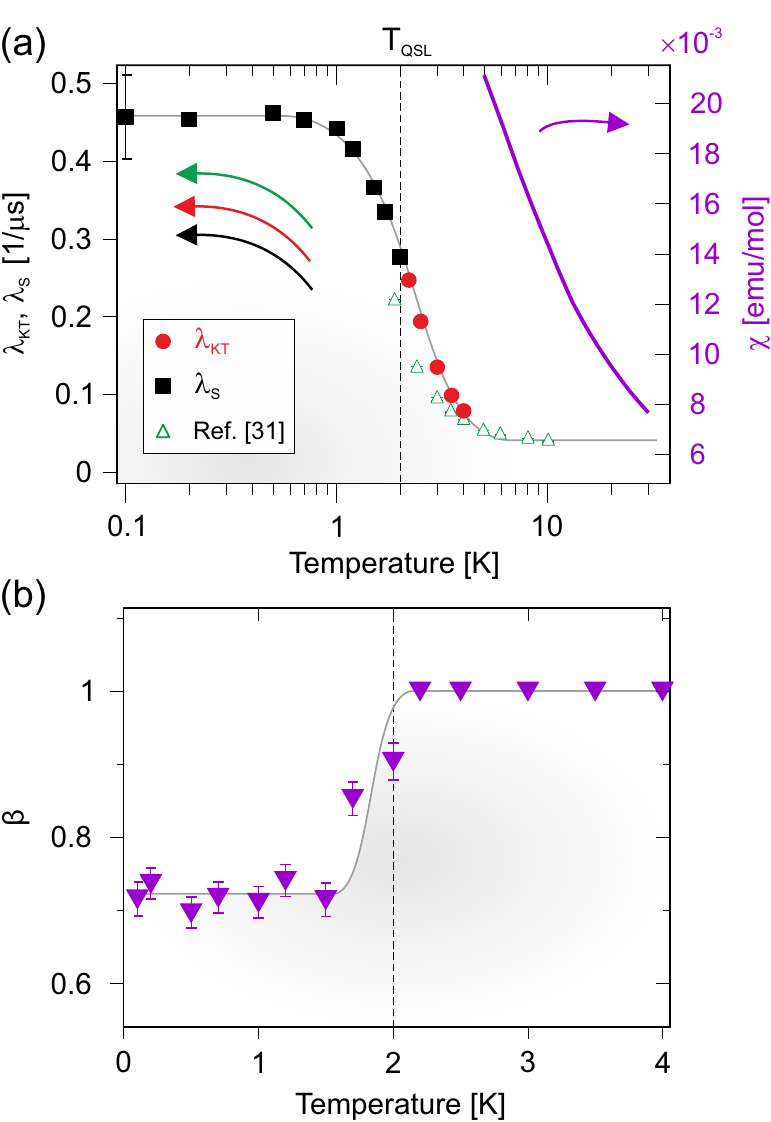}
  \end{center}
  \caption{Temperature dependencies of the obtained fit parameters using Eq.~(\ref{eq:ZF_high}) and Eq.~(\ref{eq:ZF_low}): (a) The relaxation rates ($\lambda_{\rm KT}$, $\lambda_{\rm S}$) and (b) the stretched exponent ($\beta$). For reference, the low temperature part of the magnetic susceptibility is also included, as well as the relaxation rate obtained in a previous $\mu^+$SR study at higher temperatures \cite{Chow2012}. Formation of the QSL state is indicated by the vertical dashed line, $T_{\rm QSL}\approx2$~K.}
  \label{fig:ZFPara}
\end{figure}

The temperature dependencies of the obtained ZF fit parameters are displayed in Fig.~\ref{fig:ZFPara}. In the whole temperature range, $A_{\rm KT}\simeq0.09\simeq A_{\rm S}$ was obtained, signifying the quality of the fits. The relaxation rates, $\lambda_{\rm S}$ and $\lambda_{\rm KT}$, are shown together with the measured magnetic susceptibility (down to 5~K). Indeed, the relaxation rates follow the measured $\chi$ meaning the tail of $\chi$ (Fig.~\ref{fig:Crystal}(b)) observed at lower temperature is a result of imperfect spin singlet formation. These unpaired spins are responsible for the fluctuations still present and thus the increase in $\lambda_{\rm KT}$ and $\chi$. In other words, the tail feature observed in $\chi$ is not due to impurities as previously speculated \cite{Isobe2006}. Below 2~K, the fit favors $\Delta_{\rm KT}=0$~$\mu$s$^{-1}$ and Eq.~(\ref{eq:ZF_high}) can no longer reproduce the data, which is instead replaced by Eq.~(\ref{eq:ZF_low}). The crossover between the two regions has no impact on the fitting results as smooth transition from $\lambda_{\rm KT}$ to $\lambda_{\rm S}$ is observed, suggesting the assessments of the fit functions to be correct \cite{Fit_Note}. Unexpectedly, a temperature independent behavior of $\lambda_{\rm S}$ is observed at lower temperatures [Fig.~\ref{fig:ZFPara}(a)]. This would suggest that the fluctuations are no longer thermally driven, i.e. below $T_{\rm QSL}\approx2$~K, quantum fluctuations are instead dominating. Similar and supporting trend is found for the temperature dependence of the stretched exponent parameter ($\beta$), which poses values close to 1 at 2~K and slowly decrease and saturate to a value of $\beta\approx0.7$. The value of 1 is naturally expected at higher temperature. As thermal fluctuations is suppressed, $\beta$ takes on values $<1$. The quantum fluctuations present at low temperatures, which by all means are uncorrelated, result in a distribution of spin-spin correlation times. This kind of distribution forces $\beta<1$. While it is difficult to assert any details from the specific value of $\beta\approx0.7$, a clear decrease from 1 is none the less observed. In the literature, a change in $\beta$ together with temperature independent relaxation rate has been observed in other QSL compounds, such as the 1D chain K$_3$Cu$_3$AlO$_2$(SO$_4$)$_4$ \cite{Fujihala2017}, the 2D triangular antiferromagnet YbMgGaO$_4$ \cite{Yuesheng2016} or the 3D antiferromagnet PbCuTe$_2$O$_6$ \cite{Khuntia2016}. 
 
%NaYbO$_2$ \cite{Ding2019} 
 
\begin{figure}[ht]
  \begin{center}
    \includegraphics[keepaspectratio=true,width = 80 mm]{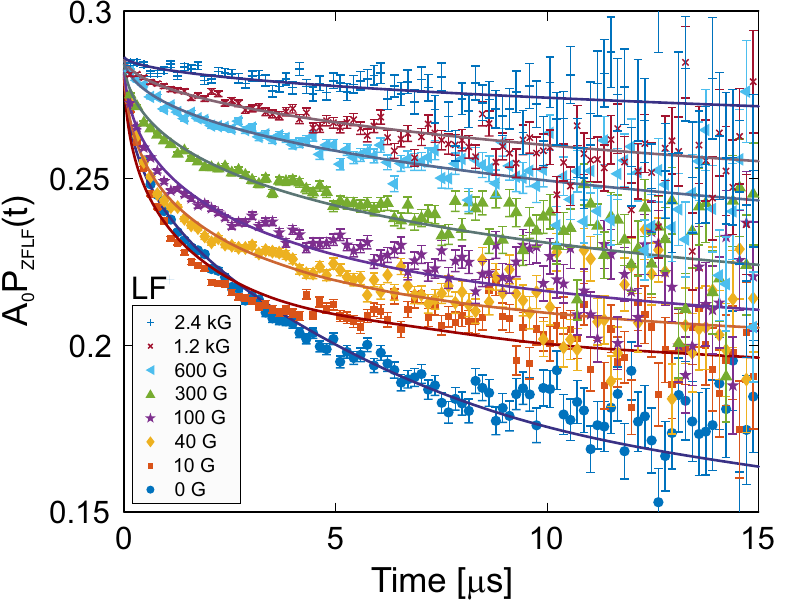}
  \end{center}
  \caption{Longitudinal field (LF) $\mu^+$SR spectra at selected LF collected at $T=100$~mK. The solid lines represents fit with Eq.~(\ref{eq:ZF_low}) using a global fit procedure described in the text.}
  \label{fig:LFSpec}
\end{figure}

It should of course be noted that static fields may in certain cases cause exponential like relaxation. In order to confirm the dynamical origin of the relaxation rate, the sample was measured under a series of longitudinal fields (LF), as shown in Fig.~\ref{fig:LFSpec}. The initial decoupling observed at LF~=~10~G comes from the background signal along with a very small impurity fraction (asymmetry~$\sim$~0.01). The latter was also confirmed in the basic XRD characterisation performed after sample synthesis. Such very minor contribution has still been readily accounted for in the analysis and could be subtracted without any influence on the final result. What is very clear is that the spectra are not fully decoupled even at LF~=~2400~G (Fig.~\ref{fig:LFSpec}), suggesting that the ground state is highly dynamic. The LF spectra were fitted in accordance to the ZF model, $i.e.$ Eq.~(\ref{eq:ZF_low}). Although, to reduce the number of parameters, $A_{\rm S}$ was kept as a global (free) parameter, while $A_{\rm BG}$ and and $\beta$ were fixed to the values as previously determined.

%A discepency is may be present at 10~G because of the minor impunity decoupling that occurs. 

\begin{figure}[ht]
  \begin{center}
    \includegraphics[keepaspectratio=true,width = 75 mm]{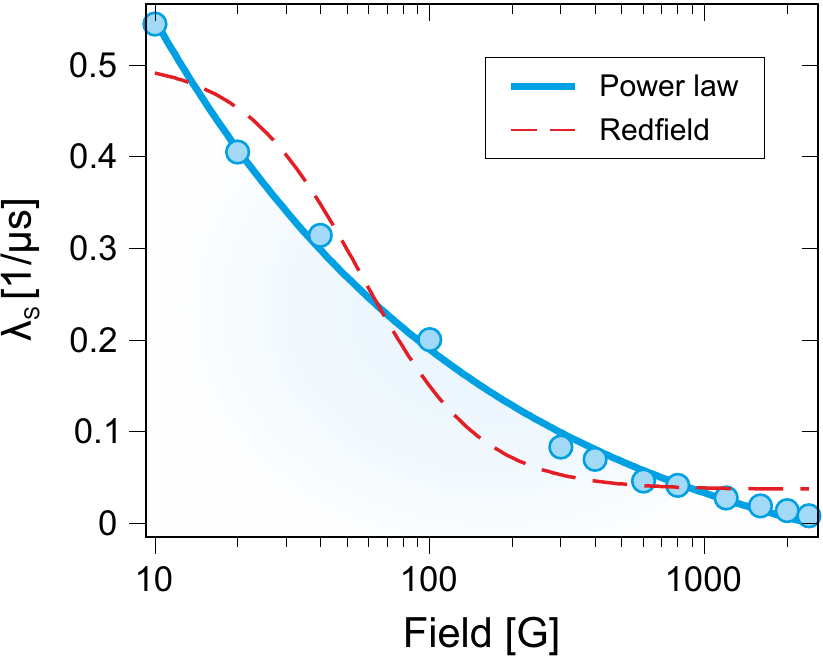}
  \end{center}
  \caption{Relaxation rate ($\lambda_{\rm S}$) as a function of externally applied LF plotted in logarithmic scale. The solid blue line represents a fit using a power law: $\lambda_{\rm S}=\frac{1}{1+cB^\alpha}$ with $\alpha=0.742(2)$, whereas the dashed red line is a fit with the Redfield equation: $\lambda_{\rm S}(B)=\frac{2\gamma_{\mu}^2\Delta^2\tau}{1+(\gamma_{\mu}B\tau)^2}$, where $\gamma_{\mu}$ is the muon gyromagnetic ratio, $\Delta$ is the internal field distribution width and $\tau$ is the spin-spin correlation time.}
  \label{fig:LFPara}
\end{figure}

Consequently, the fit procedure effectively reduced the number of free and field-dependent fit parameters to one, the relaxation rate ($\lambda_{\rm S}$). This parameters is plotted as a function of the applied LF in Fig.~\ref{fig:LFPara}. At lower fields, a value close to the ones obtained at ZF is found but a decreases as a function of LF is observed. Since Eq.~(\ref{eq:ZF_low}) is a stretched exponential, quantitative statement about the absolute values are difficult to justify. However, the information from the relative field dependence can still be extracted. In the motional narrowing limit, the relaxation rate is expected to follow the field dependence of Fourier transform of the dynamical correlation function, commonly known as the Redfield equation: $\lambda_{\rm S}(B)=\frac{2\gamma_{\mu}^2\Delta^2\tau}{1+(\gamma_{\mu}B\tau)^2}$. A fit with the Redfield function does not yield a good representation of the field-dependence $\lambda_{\rm S}(LF)$ (see Fig.~\ref{fig:LFPara} for more details). Therefore, the data could instead be very well fitted using a more general power law form: $\lambda_{\rm S}=\frac{1}{1+cB^\alpha}$ for which $\alpha=0.742(2)$ was obtained. It is noted that LF relaxation rate maps the Fourier transform of the dynamical correlation function, meaning the field dependence of $\lambda_{\rm S}$ (Fig.~\ref{fig:LFPara}) corresponds to the spin-spin correlation function. For a gapped spin liquid, the correlation function is expected to experience a exponential like decay (i.e. Redfield like) \cite{Schollwock2008}. On the other hand, for a gapless spin liquid, $e.g.$ Tomonaga-Luttinger liquid (TLL), the correlation is expected to show a power law behavior \cite{Schollwock2008}. Based on the fits presented in Fig.~\ref{fig:LFPara}, we therefore initially suggest that the QSL ground state in K$_2$V$_8$O$_{16}$ could be the gapless TLL. 

It is worth noticing that the signal from quantum fluctuations originating from the spin singlet dimers would result in a gaped spin liquid. Naturally, such fluctuations would also contribute to the LF relaxation rate and will be included into the $\alpha$ coefficient of the power law (Fig.~\ref{fig:LFPara}). However, the obtained $\alpha=0.742(2)$ is a strong deviation from the ideal Redfield ($\alpha=2$), which supports a 1D diffusive picture of the spin fluctuations. We anticipate that the non-gapped nature (TLL) will be directly confirmed via neutron scattering at future high-intensity/-resolution neutron spallation sources.

%{\color{red} Of course, quantum fluctuations originating from the spin singlet dimers would result in a gaped spin liquid. Naturally, these fluctuations contribute to the LF relaxation rate as well and such information is most likely included in the obtained coefficient $\alpha=0.742(2)$. However, a strong deviation from 2 is observed (Redfield) and supports a 1D diffusive picture of the spin fluctuations. Regardless though, the main message, formation of QSL is none the less confirmed. We hope the gaped nature can be directly confirmed via neutron scattering in future neutron spallation sources.}

%The relaxation rate is the Fourier transform of the dynamical correlation function, the field dependence corresponds to the spin-spin correlation function. For a gapped spin liquid, the correlation function is expected to experience a exponential like decay. A none gapped spin liqudi on the other hand, e.g. Tomonaga-Luttinger liquid, the correlation is expceted to show a power law behaviout. A fit utilising an expnential, the traditional refield and a powerlaw behavior is shown in \ref{fig:ZFPara}. It is noted that excluding the 10G measuremetn does not alte the results and excluding low field values are common in muSR in order to properly fit the region without contribution from nuclear moments. Clearly the behavior is a power law decrease, suggesting it may be a Tomonaga luttinger spin liquid. 

While there are several types of QSL, let us focus on the TLL, which is a liquid state present in one dimensional heavily correlated electron systems. In fact, K$_2$V$_8$O$_{16}$ may be checking these boxes. DFT+U calculations \cite{Kim2016} show a highly one dimensional character with a flat Fermi surface. It was proposed that local electron correlation is driving the charge ordering and orbital ordering of the V t$_{2g}$ bands. Therefore, K$_2$V$_8$O$_{16}$ is both highly correlated and 1D. An important consequence of TLL is the prediction of spin-charge separation. Typical problem with experimental verification of TLL is the fact that many QSL are usually stabilised under extreme conditions ($i.e.$ high pressure or magnetic field e.g. BaCo$_2$V$_2$O$_8$ \cite{BCVO_1,BCVO_2}). In K$_2$V$_8$O$_{16}$ however, the QSL/TLL state is stabilized at ambient condition (zero pressure and magnetic field), although at very low temperatures. Hence, one would expect a FM or AF ordered states to be stabilized either at high magnetic field or pressure, at least according to the phase diagram of the XXZ Hamiltonian \cite{Schollwock2008}. Verifying these quantities in future experiments would clearly strengthen the TLL scenario for K$_2$V$_8$O$_{16}$.
%As a matter of fact, it was recently shown that hydrostatic pressure stabilises a dynamic AF order in K$_2$V$_8$O$_{16}$. 
%and Rb$_2$V$_8$O$_{16}$ is an insulating short range partially ordered state at 1.6~K \cite{Forslund2021_Rb}

Let us take a step back and compare to other members in the Hollandite family. K$_2$Ti$_8$O$_{16}$ is a Pauli paramagnetic metal down to the lowest measured \cite{Isobe2009} while K$_2$Cr$_8$O$_{16}$ is a FM insulator down to 2~K \cite{Hasegawa2009}. Similarly, Rb$_2$Cr$_8$O$_{16}$ is a FM metal \cite{Sugiyama2012}. While it is clear that the Hollandite family exhibits a variety of ground states, a QSL was up until now never reported within the family, even though band structure calculations \cite{Toriyama2011_Ru} of K$_2$Ru$_8$O$_{16}$ in fact proposed it to belong to the class of TLL. Hence, this work highlights K$_2$V$_8$O$_{16}$ as the first Hollandite to exhibit QSL (possibly TLL) behavior, which originates from the high-temperature CO formation of isolated 1D spin-$1/2$ Heisenberg chains. It is well established that CO can play an important role in order to stabilise various phases. It was shown to be the driver from low to high Spin-State transition in the AF YBaCo$_2$O$_5$ \cite{Vogt2000} and multiferroicity is naturally heavily linked to charge ordering \cite{Jeroen2008, Yamauchi2010}. In fact, detailed CO engineering was proposed in order to induce desired multiferroic properties \cite{He2016}. Likewise, CO plays an important part in high temperature superconductivity. CO appears usually in the vicinity of superconducting state to either co-exist or to compete with each other \cite{Tranquada1997}. Therefore, controlling the CO are believed to be a key component in order to achieve and fully understand the paring mechanisms in both the archetypal cuprate superconductor YBa$_2$Cu$_3$O$_{7-\delta}$ \cite{YBCO_CO_1, YBCO_CO_3} as well as others \cite{Eduardo2014,YBCO_CO_2,Stripe_CO}. In a similar fashion, K$_2$V$_8$O$_{16}$ is an example on how the formation of QSL is stabilized by the CO formation, which effectively reduces the dimensionality to a true 1D system within the 3D crystallographic structure. To the best of our knowledge, this is the first clear instance of a CO induced QSL state.

%For the sake of comlpetness, we include a scenario which in cludes short range ordering. From $\mu^+$SR, a short range ordering would develope a relacation similar to a streched exponential. Short range order has been observed in many systems, especially for low dimensional systems like NaNiO$_2$ \cite{Baker2005, Forslund2020_Na}, LaCo$_2$P$_2$ \cite{Forslund2021_La}, CrCl3 \cite{Forslund2022} or BaCo$_2$V$_2$O$_8$ \cite{Mansson2012}. However, the temperature independent behabiour would be difficult to explain with SRO. 

%\section{\label{sec:conclusion}Conclusions}
In summary, our $\mu^+$SR establish the absence of long- or short-range magnetic order in K$_2$V$_8$O$_{16}$ down to $T=100$~mK. The ground state is found to be dynamic and temperature independent, which matches the properties of a quantum spin liquid (QSL) forming below $T_{\rm QSL}\approx2$~K. Field dependent data confirms the dynamic nature and points towards a Tomonaga-Luttinger liquid (TLL) phase, which uniquely forms in this stoichiometric compound under zero field and ambient pressure. The key to such properties is the formation of charge order (CO) within half of the quasi-one-dimensional chains, which was confirmed by XRD \cite{Komarek2011}. Combined with theoretical predictions of spin-singlet formation \cite{Kim2016} along with supporting experimental evidence of a strong decrease in magnetic susceptibility [Fig.~\ref{fig:Crystal}(b)]] suggest a mechanism that effectively isolates the remaining spin chains and thereby stabilizes a QSL. These findings highlights the high-pressure synthesis route as a platform for discovering emerging materials and novel physical properties. Finally, we also establish the hollandites as a new group of model compounds available to study intrinsic QSL behavior as well as its response to future external tuning via magnetic field, high pressure and chemical substitution/doping.

%the absence of any oscillatory signal clearly excludes any long range ordering. The temperature independent dynamics present below 2~K suggest the compound to be a a quantum spin liquid (QSL). Being the first in the family to exhibit such characteristics, K$_2$V$_8$O$_{16}$ represent a unique case of QSL, where the higher temperature charge ordering (CO) effectively transformed the compound into a model Heisenberg spin-$1/2$ chain. The presence of electron correlation and the highly 1D nature would suggest K$_2$V$_8$O$_{16}$ to be a Tomonaga-Luttinger liquid (TLL), which we hope future experimental work could confirm. While the size of the crystals limits the available experimental techniques, we have manage to assert the ground state based on powder sample. 

%In detail, absence of long range magnetic order is confirmed by the exponential like relaxation at temperatures down to $T=100$~mK. Moreover, longitudinal field (LF) measurements confirm the origin of these temperature independent relaxation to be dynamic and supports the scenario of a QSL. These results represent the first instance of a QSL in the hollandite family, which together with the 1D nature and electron correlation present makes K$_2$V$_8$O$_{16}$ a prime candidate for a Tomonaga-Luttinger liquid (TLL). 

\begin{acknowledgments} 
This research was supported by the Swedish Research Council (VR) via a Neutron Project Grant (Dnr. 2016-06955) and the Carl Tryggers Foundation for Scientific Research (CTS-18:272). E.N. is funded by the Swedish Foundation for Strategic Research (SSF) within the Swedish national graduate school in neutron scattering (SwedNess). D.A. acknowledges financial support from the Romanian UEFISCDI project PN-III-P4-ID-PCCF-2016-0112, Contract Nr. 6/2018. Y.S. is funded by VR through a Starting Grant (Dnr. 2017-05078) as well as the Chalmers Area of Advance-Materials Science. J.S. acknowledges support from Japan Society for the Promotion Science (JSPS) KAKENHI Grants No. JP18H01863 and No. JP20K21149.
%All images involving crystal structure were made with the Vesta software\cite{Momma2011} and the $\mu^+$SR data was fitted using \texttt{musrfit} \cite{musrfit}.
\end{acknowledgments}

%------------------------------------ REFRERENCES
\bibliography{Refs} % Call your Name.bib file with all the references.
\end{document}